# Using the Jupyter Notebook as a Tool for Open Science: An Empirical Study

Bernadette M. Randles
Department of Information Studies
University of California, Los Angeles
USA

Irene V. Pasquetto
Department of Information Studies
University of California, Los Angeles
USA

Milena S. Golshan
Department of Information Studies
University of California, Los Angeles
USA

Christine L. Borgman
Department of Information Studies
University of California, Los Angeles
USA

*As scientific work becomes more computational and data-intensive, research processes and results become more difficult to interpret and reproduce. In this poster, we show how the Jupyter notebook, a tool originally designed as a free version of Mathematica notebooks, has evolved to become a robust tool for scientists to share code, associated computation, and documentation.*

## I. INTRODUCTION AND BACKGROUND

The Jupyter notebook is an open-source, browser-based tool functioning as a virtual lab notebook to support workflows, code, data, and visualizations detailing the research process. It is machine and human-readable, which facilitates interoperability and scholarly communication. These notebooks can live in online repositories and provide connections to research objects such as datasets, code, methods documents, workflows, and publications that reside elsewhere. Jupyter notebooks are one means to make science more open. Their relevance to the JCDL community lies in their interaction with multiple components of digital library infrastructure such as digital identifiers, persistence mechanisms, version control, datasets, documentation, software, and publications. Our poster examines how Jupyter notebooks embody the FAIR (Findable, Accessible, Interoperable, Reusable) principles for digital objects and assess their utility as viable tools for scholarly communication [11].

In our preliminary work, we investigated how scientists have begun to cite Jupyter notebooks in their published papers in astronomy and related fields. The study is based in astronomy because it is a field that has made great progress toward open science. In recent years, the ability to interpret findings has come to depend on access to data and to the procedures on which research is based. The desire to inspect data is among the drivers for requiring researchers to share data at the time of paper publication. Funding agencies and journals alike commonly require data release as a condition for funding or publication. Other arguments for data sharing include reproducibility, reuse, and transparency of research processes [3,4]. However, sharing data alone is rarely sufficient to interpret findings that result from complex computational processes. Also necessary is access to the software that generated data, research workflows, instrumentation details, laboratory notebooks (digital or otherwise), and other documentation that constitutes the "recipe" for how the research was accomplished. Jupyter notebooks are growing in popularity as a tool to integrate these resources and to provide access to them.

The FAIR principles are a promising new development to operationalize the notion of open science. FAIR combines open access to data, software, and other components of the research process by encompassing all "research objects" associated with a research project or paper. The idea of a "research object" or "scholarly research object" emerged from discussions of scholarly communication, research tools, and knowledge infrastructures to address eScience and other changes in the nature of scholarly communication [1,2,5]. As implemented in the FAIR principles, a research object can be either the entirety of the set of data and tools involved in an act of scientific discovery or a subset of a larger set of research objects. Reusability is a nebulous term, containing within it the concepts of reproducibility and replicability. For this poster, we use Peng's definitions of *computational reproducibility* as the availability of the data and code used in the computation and *replicability* as the ability to obtain the same results on data collected independently [8].

## II. METHODS

Drawing a complete list of all mentions of Jupyter notebooks in the Astrophysics Data System, we identified 91 relevant publications. The Astrophysics Data System, developed by NASA, is a comprehensive catalog of papers in astronomy and physics. Each paper (article, report, or preprint) that mentioned a Jupyter notebook was coded for how the uses contributed to open science. We coded for mentions, citations, or other explicit linkages to specific Jupyter notebooks and to research objects such as datasets, code, and publications.

## III. RESULTS

We identified eight ways in which scientists mentioned the Jupyter notebooks and five ways in which they provided access, only some of which appeared to facilitate open science. Of the 91 papers, 37 linked to openly accessible Jupyter notebooks containing detailed research procedures, associated code, analytical methods, and results. Another 54 papers mentioned a Jupyter notebook in the text, but did not provide access to one. Practices for mentioning, storing, and providing access to the notebooks varied greatly across papers.

In our poster we demonstrate that authors use Jupyter notebooks for a variety of tasks in the research process ranging from data construction to analysis or the manuscript itself. Thirty-seven papers provided explicit descriptions of how they used a notebook, and a few provided only general descriptions of usage. In some cases, researchers appeared to be evaluating Jupyter notebooks as research tools. Twenty-three mentioned the iPython or Jupyter projects by reference to one or more of the papers by Perez and Granger [9], 21 referenced the main Jupyter website, and one paper mentioned a Jupyter notebook that required permissions for access. Most papers mention the notebooks in multiple ways; the most popular method was via a link to NBviewer or to a GitHub repository.

We present one case scenario that explicitly mentioned open science goals as a means to demonstrate roles that Jupyter notebooks can play in making science more open. To demonstrate how this paper implemented the FAIR guidelines, we identified the specific mentions and followed the hyperlinks to the associated Jupyter notebooks. We outline linkages to other objects and processes, with a focus on discoverability. By tracing the ways in which this paper documents and links to research objects, we identify ways in which the authors are using Jupyter notebooks to make their own science more open.

The scenario paper is an astronomy article examining masses of binary stars, drawing on data from several telescopes [6]. In the conclusion, the authors state, "In the effort of open and reproducible research, we have made several data products freely available." The paper links to data repositories in Zenodo and GitHub. Their GitHub repository contains Jupyter notebooks, .csv files of associated data, and python scripts. These researchers are using the Jupyter notebooks to release data and code by providing context for many of the research objects integral to the research results reported in the paper. They provided the code enumerating the procedures and processes to refine the data. Their Jupyter notebooks contain steps and code (or function calls to code) with explanations of refinement and analysis procedures. Their python scripts, submitted paper, and related data are available in the GitHub repository. In sum, the authors of the scenario paper applied the FAIR principles by providing open access to all research objects included in their study.

## IV. CONCLUSIONS

We conclude our poster with questions for future work: What information should an "ideal" Jupyter notebook contain to facilitate the FAIR principles most effectively? How can or should software and other tools be stabilized or preserved, when access is provided via published papers? How can persistence mechanisms such as Object Reuse and Exchange or ResourceSync be used to facilitate the stability of Jupyter notebooks [7,10]?

While much is known about the incentives and disincentives for sharing data [3], little is yet known about motivations for uses of specific tools such as Jupyter notebooks for making science more open. The Jupyter notebook is a promising tool for open science. These notebooks are part of a constellation of open science activities, such as the open repositories within which they reside. The larger problems of sustaining access to research objects and the relationships between them are a continuing challenge. Meanwhile, studying components of open science may lead to answers to some of these larger questions.

## ACKNOWLEDGMENT

This research was supported by a grant from the Alfred P. Sloan Foundation (#201514001, C.L. Borgman, PI). We are grateful to Fernando Perez for discussions about the origins and goals of Jupyter Notebooks. We also thank Peter T. Darch for commenting on earlier drafts of this paper.